\documentclass[aip,jcp]{revtex4}
\pdfoutput=1

\usepackage{amsbsy}
\usepackage{amsmath}
\usepackage{dcolumn}
\usepackage{graphicx}

\begin{document}

\title{Path-integral calculation of the third virial coefficient of quantum
  gases at low temperatures
\footnote{Partial contribution of the National Institute of Standards and
  Technology, not subject to copyright in the United States.}
}
\author{Giovanni Garberoglio}
\affiliation{European Centre for Theoretical Studies in Nuclear Physics and
  Related Areas (FBK-ECT*), via delle Tabarelle 256, I-38122 Villazzano
  (TN), Italy}
\altaffiliation{Formerly at: Interdisciplinary Laboratory for Computational Science (LISC),
  FBK-CMM and University of Trento, via Sommarive 18, I-38123 Povo (TN),
  Italy.}
\email[Corresponding author. Email: ]{garberoglio@ectstar.eu}
\author{Allan H. Harvey}
\affiliation{Thermophysical Properties Division, National Institute of
  Standards and Technology, 325 Broadway, Boulder CO 80305, USA.}

\date{\today}

\begin{abstract}
We derive path-integral expressions for the second and third virial
coefficients of monatomic quantum gases.  Unlike previous work that
considered only Boltzmann statistics, we include exchange effects
(Bose--Einstein or Fermi--Dirac statistics).
We use state-of-the-art pair and three-body potentials to calculate
the third virial coefficient of ${}^3$He and ${}^4$He 
in the temperature range $2.6 - 24.5561$~K.
We obtain uncertainties smaller than those of the limited experimental
data.  Inclusion of exchange effects is necessary to obtain accurate
results below about 7~K.
\end{abstract}

\maketitle

\section{Introduction}

Thermodynamic properties of fluids at very low temperatures are of
significant interest.  For example, the current International
Temperature Scale~\cite{its90} makes use of volumetric properties and
vapor pressures of helium isotopes below the triple point of neon
(24.5561~K); below the triple point of hydrogen (13.8033~K), the scale
is based entirely on properties of ${}^3$He and ${}^4$He.  The
theoretical analysis of relevant properties at these conditions, such
as the virial coefficients that describe the fluid's departure from
ideal-gas behavior, is complicated by the presence of quantum effects.

The inclusion of quantum effects in the calculation of virial
coefficients was one of the first numerical applications of the
Path-Integral Monte Carlo (PIMC) method.~\cite{FH} In a series of
pioneering works published in the 1960's, Fosdick and Jordan
showed how to calculate the second and third virial coefficient of a
monatomic gas using computer
simulations.~\cite{fosdick-jordan66,jordan-fosdick68,fosdick68} Given
the limited computational resources available at that time, they were
able to calculate the third virial coefficient only
in the case of two-body interactions, using a model potential of the
Lennard-Jones form and assuming distinguishable particles (Boltzmann
statistics). They argued that their method could be extended to
include the proper quantum statistics, but they were able to compute
exchange effects only in the case of the second virial coefficient.

Recently, the exponential increase in computational power 
has enabled use of the path-integral method to
calculate the properties of quantum degenerate systems, notably
superfluid helium.~\cite{Ceperley}
At the same time, progress in the computation of {\em ab initio}
electronic properties of interacting atoms resulted in the availability of
very precise two- and three-body interparticle potentials, at least for
the lightest particles such as helium
atoms~\cite{Jez07,HBV07,He-3body,CPS09,He2-2010} or hydrogen
molecules.~\cite{diep-johnson00,virial-H2}

A natural application for these potentials is the calculation of
virial coefficients. As is well known, the second virial coefficient
depends only on the two-body potential, the third virial coefficient
depends only on two-body and three-body interactions, etc. The second
virial coefficient for a monatomic gas can be rigorously obtained at
the fully quantum level from the calculation of the phase shifts due
to the pair potential, and previous work has shown that a completely
{\em ab initio} calculation of second virial coefficients for helium
can have uncertainties comparable to and in many cases smaller than
those of the most precise
experiments.~\cite{phi07,hur00,BHV07,Mehl09,Cencek11}

In the case of the third virial coefficient, no closed-form solution
of the quantum statistical mechanics problem is known. First-order
semiclassical approaches have been derived~\cite{yokota60,ram-singh73}
and show that, in the case of helium, quantum diffraction effects
result in significant modifications of the classical result, even at
room temperature. However, there is no rigorous way to evaluate the
accuracy or uncertainty of the semiclassical result, especially at low
temperatures. 

In recent work,~\cite{Garberoglio2009b} we extended the methodology
pioneered by Fosdick and Jordan, deriving a set of formulae allowing a
path-integral calculation of the third virial coefficient $C(T)$ of
monatomic species for arbitrary two- and three-body potentials. Our
results were limited to Boltzmann statistics (i.e., distinguishable
particles) and we did not present results for temperatures lower than
the triple point of neon ($24.5561$~K), which we deemed to be a
reasonable lower bound so that exchange effects could be neglected.
Nevertheless, we were able to compute the value of the third virial
coefficient of ${}^4$He with an uncertainty one order of magnitude
smaller than that of the best experiments.

Recent experimental results overlapping with our temperature
range,~\cite{Gaiser09,Gaiser10} although mostly consistent with our
calculations, seemed to indicate a systematic deviation which the
authors speculated could originate from our neglect of the proper
quantum statistics of helium atoms.

In this paper, we extend our computational methodology to calculate
the quantum statistical contributions to the third virial coefficient,
and compute $C(T)$ for both isotopes of helium in the temperature
range $2.6 - 24.5561$~K, extending the temperature range considered in
our previous work down into the range where exchange effects are
important.
We show that quantum statistical effects are significant only for
temperatures smaller than about $7$~K, and compare our results to
low-temperature experimental data.

In a subsequent publication~\cite{paper2}, we will present results
covering the entire temperature range (improving on our previous
results for ${}^4$He at 24.5561~K and above) with rigorously derived
uncertainties.  We will also extend our methodology to include
acoustic virial coefficients, and compare those calculations to
available data.  In the present work, our focus is on low temperatures
and specifically on the effect of non-Boltzmann statistics.

\section{Path-integral calculation of the virial coefficients}

The second and third virial coefficients, $B(T)$ and $C(T)$ respectively, are
given by~\cite{hcb}
\begin{eqnarray}
  B(T) &=& -\frac{1}{2 V} \left( Z_2 - Z_1^2 \right) \label{eq:B} \\
  C(T) &=& 4 B^2(T) - \frac{1}{3 V} \left[ Z_3 - 3 Z_2 Z_1 + 2 Z_1^3
  \right], \label{eq:C}
\end{eqnarray}
where $V$ is the integration volume (with the limit $V
\rightarrow \infty$ taken at the end of the calculations), and the functions
$Z_N$ are given by:
\begin{eqnarray}
Z_3 &=& \Lambda^9 \int \mathrm d1 \mathrm d2 \mathrm d3 ~ 
\langle 1 2 3 | \mathrm e^{-\beta \hat H_3} \sum_{\pi_3} {\cal P}_{\pi_3} | 1 2 3
\rangle \label{eq:Z3} \\ 
Z_2 &=& \Lambda^6 \int \mathrm d1 \mathrm d2 ~
\langle 1 2 | \mathrm e^{-\beta \hat H_2 } \sum_{\pi_2} {\cal P}_{\pi_2} | 1 2 \rangle
 \label{eq:Z2} \\ 
Z_1 &=& \Lambda^3 \int \mathrm d1 ~
\langle 1 | \mathrm e^{-\beta \hat H_1} | 1 \rangle  = V, \label{eq:Z1}
\end{eqnarray}
where $\hat H_N$ is the $N$-body Hamiltonian, $\beta = 1/(k_\mathrm B
T)$, ${\cal P}$ is a permutation operator (multiplied by the sign of
the permutation in the case of Fermi--Dirac statistics), the index
$\pi_3$ runs over the 6 permutations of 3 objects (i.e., $123$, $132$,
$213$, $321$, $231$ and $312$), and $\pi_2$ runs over the 2
permutations of 2 objects (i.e., $12$ and $21$).  $\Lambda = h /
\sqrt{2 \pi m k_\mathrm B T}$ is the thermal de~Broglie wavelength of
a particle of mass $m$ at temperature $T$.  For the sake of
conciseness, we denote by $| i \rangle$ an eigenvector of the position
operator relative to particle $i$ and by $\mathrm d i$ ($i=1,2,3$) the
integration volume relative to the Cartesian coordinates of the $i$-th
particle.  Note that, in order to produce the molar units used by
experimenters and in our subsequent comparisons with data, the right
side of Eq.~(\ref{eq:B}) and the second term in the right side of
Eq.~(\ref{eq:C}) must be multiplied by Avogadro's number and its
square, respectively.

In the following, we will derive a path-integral expression for the
calculation of the virial coefficients with Eqs.~(\ref{eq:B}) and
(\ref{eq:C}). We perform the derivation in detail in the case of
$B(T)$ to establish the notation, and then extend the results to the
more interesting case of $C(T)$.

\subsection{Second virial coefficient}

In this paper, we adopt Cartesian coordinates to describe the atomic
positions. This differs from the approach developed in
Refs.~\onlinecite{jordan-fosdick68} and \onlinecite{Garberoglio2009b},
where Jacobi coordinates were used. This choice allows the exchange
contribution to be computed in a much simpler manner than would be the
case if Jacobi coordinates were used, especially in the case of three
or more particles.

From Eqs.~(\ref{eq:B}) and (\ref{eq:Z2}), it can be seen that there
are two contributions to $B(T)$. The first one comes from considering
the identity permutation only, and takes into account only quantum
diffraction effects. This is the only contribution that gives a
nonzero result at high temperatures, where the particles can be treated
as distinguishable (Boltzmann statistics).

The second contribution to $B(T)$, which we will call {\it exchange}
(xc), comes from the only other permutation involved in the definition
of the quantity $Z_2$ above.

The expression of these two contributions in Cartesian coordinates is:
\begin{widetext}
\begin{eqnarray}
  B_\mathrm{Boltzmann}(T) &=& -\frac{\Lambda^6}{2 V} 
\int \mathrm d \boldsymbol{r}_1 \mathrm d \boldsymbol{r}_2 ~
\langle \boldsymbol{r}_1 \boldsymbol{r}_2 |
\exp\left[ -\beta (\hat K_2 + \hat U_2(|\boldsymbol{r}_2 - \boldsymbol{r}_1|)
  ) \right] - \exp\left[ -\beta \hat K_2 \right] 
| \boldsymbol{r}_1 \boldsymbol{r}_2 \rangle \label{eq:Bdir} \\
  B_\mathrm{xc}(T) &=& - \frac{(-1)^{2I}\Lambda^6}{(2I+1)2 V} 
  \int \mathrm d \boldsymbol{r}_1 \mathrm d \boldsymbol{r}_2
\langle \boldsymbol{r}_1 \boldsymbol{r}_2 |
\exp\left[ -\beta (\hat K_2 + \hat U_2(|\boldsymbol{r}_2 - \boldsymbol{r}_1|)
  ) \right]| \boldsymbol{r}_2 \boldsymbol{r}_1 \rangle,
\label{eq:Bx}  
\end{eqnarray}
\end{widetext}
where we denote by $K_N$ the total kinetic energy of $N$ bodies and by $\hat
U_2(r)$ the two-body potential energy operator and $I$ is the nuclear spin
of the atomic species under consideration ($I=0$ for ${}^4$He and $I=1/2$
for ${}^3$He). 

Equations~(\ref{eq:Bdir}) and (\ref{eq:Bx}) can be rewritten by using the
Trotter identity
\begin{equation}
  \mathrm e^{\hat K_2 + \hat U_2} = \lim_{P \rightarrow \infty} \left( \mathrm e^{\hat
    K_2/P} \mathrm e^{\hat U_2/P} \right)^P
\label{eq:Trotter}
\end{equation}
with a positive integer value of the Trotter index $P$.

Following the procedure outlined in Ref.~\onlinecite{Garberoglio2009b}, one can
then write $B_\mathrm{Boltzmann}(T)$ as
\begin{eqnarray}
  B_\mathrm{Boltzmann}(T) = - 2 \pi^2 \int_0^\infty r^2 \mathrm dr ~ 
\left(\exp\left[-\beta \overline{U}_2(r)\right] - 1 \right),
\label{eq:BBoltz}
\end{eqnarray}
where the two-body effective potential $\overline U_2(r)$ is given by
\begin{widetext}
\begin{eqnarray}
  \exp\left[-\beta \overline{U}_2(r)\right]  &=& 
\int \prod_{i=1}^{P-1} \mathrm d\Delta\boldsymbol{x}_1^{(i)} 
\mathrm d\Delta\boldsymbol{x}_2^{(i)} ~ 
\exp\left[
-\frac{\beta}{P} \sum_{i=1}^{P} U_2(|\boldsymbol{r} + \boldsymbol{x}_2^{(i)} -
\boldsymbol{x}_1^{(i)}|) \right]  \times \nonumber \\
& &
F_\mathrm{ring}(\Delta\boldsymbol{x}_1^{(1)},\ldots,\Delta\boldsymbol{x}_1^{(P)})
F_\mathrm{ring}(\Delta\boldsymbol{x}_2^{(1)},\ldots,\Delta\boldsymbol{x}_2^{(P)})
\label{eq:U2_eff}
\\
&\underset{P \rightarrow \infty}{=}&  
  \oint {\cal D} \boldsymbol x_1  {\cal D} \boldsymbol x_2 ~ 
  \exp\left[ -\frac{1}{\hbar} \int_0^{\beta \hbar}
  \frac{m}{2}  \left| \frac{\mathrm d \boldsymbol x_1(\tau)}{\mathrm d \tau}
  \right|^2 +
  \frac{m}{2}  \left| \frac{\mathrm d \boldsymbol x_2(\tau)}{\mathrm d \tau} \right|^2 ~ 
  + U_2(|\boldsymbol r + \boldsymbol x_1(\tau) - \boldsymbol x_2(\tau)|)
  \mathrm d \tau \right],
\label{eq:U2_pi}
\end{eqnarray}  
\end{widetext}
where
\begin{equation}
F_\mathrm{ring}
= 
\Lambda^3 \left( \frac{P^{3/2}}{\Lambda^3} \right)^P 
\exp\left[ -\frac{\pi P}{\Lambda^2} \sum_{i=1}^P
\left| \Delta\boldsymbol{x}_1^{(i)} \right|^2 \right].
\label{eq:Fring}
\end{equation}

In the previous equations, we have defined
$\Delta\boldsymbol{x}_k^{(i)} = \boldsymbol{r}_k^{(i+1)} -
\boldsymbol{r}_k^{(i)}$, where $\boldsymbol{r}_k^{(i)}$ is the
coordinate of particle $k$ ($k=1,2$) in the $i$-th ``imaginary time
slice''. These ``slices'' are obtained by inserting $P$ completeness
relations of the form
\begin{equation}
  1 = \int \mathrm d\boldsymbol{r}_1^{(i)}  \mathrm d\boldsymbol{r}_2^{(i)} ~
  | \boldsymbol{r}_1^{(i)} \boldsymbol{r}_2^{(i)} \rangle \langle
  \boldsymbol{r}_1^{(i)} \boldsymbol{r}_2^{(i)} | 
\end{equation}
between the factors $\mathrm e^{\hat K_2/P}$ and $\mathrm e^{\hat U_2/P}$ of
the Trotter expansion of Eq.~(\ref{eq:Trotter}).  We used the overall
translation invariance of the system to remove the factor $V$ in
Eq.~(\ref{eq:Bdir}) and fix the $\tau=0$ slice of particle 2 at the
origin of the coordinate system. We also denoted by
$\boldsymbol{x}_1^{(i)}$ and $\boldsymbol{x}_2^{(i)}$ the coordinates
of two ring polymers having one of their endpoints fixed at the origin
($\boldsymbol{x}_1^{(1)} = \boldsymbol{x}_2^{(1)} = \mathbf 0$), and
we introduced the variable $\boldsymbol{r}$ denoting the distance
between the $\tau = 0$ time slice of the two ring polymers. In the
classical limit, where the paths $\boldsymbol x_1(\tau)$ and $\boldsymbol
x_2(\tau)$ shrink to a point, the coordinate $r$ reduces to the distance
between the particles and one has $\overline U_2(r) = U_2(r)$.

Note that the effect of the identity permutation is to set
$\boldsymbol{r}_k^{(P+1)} = \boldsymbol{r}_k^{(1)}$.  The path-integral
formalism allows one to map the quantum statistical properties of a
system with $N$ distinguishable particles (Boltzmann statistics) onto
the classical statistical properties of a system of $N$ ring polymers,
each having $P$ beads (sometimes called imaginary-time slices), which
are distributed according to the function $F_\mathrm{ring}$ of
Eq.~(\ref{eq:Fring}).~\cite{boltzmann-bias} The mapping is exact in
the $P \rightarrow \infty$ limit, although convergence is usually
reached with a finite (albeit large) value of $P$.

In the calculation of the second virial coefficient,
Eq.~(\ref{eq:BBoltz}) shows that the second virial coefficient at the
level of Boltzmann statistics is obtained from an expression similar
to that for the classical second virial coefficient, using an
effective two-body potential. This effective potential,
$\overline{U}_2(r)$, is obtained by averaging the intermolecular
potential $U_2(r)$ over the coordinates of two ring polymers,
corresponding to the two interacting particles entering the definition
of $B(T)$.

Equation~(\ref{eq:BBoltz}) is equivalent to Eq.~(19) of
Ref.~\onlinecite{Garberoglio2009b}. The only difference is that the
current approach uses Cartesian coordinates, and therefore we are left
with an average over two ring polymers of mass $m$ instead of one ring
polymer of mass $\mu = m/2$, corresponding to the relative coordinate
of the two-particle system.  The two approaches are of course
equivalent, and in fact it can be shown that Eqs.~(\ref{eq:BBoltz})
and (\ref{eq:U2_eff}) reduce to the form derived in
Refs.~\onlinecite{fosdick-jordan66} and \onlinecite{Garberoglio2009b}.
Equation~(\ref{eq:BBoltz}) is the same expression previously
derived by Diep and Johnson for spherically symmetric potentials on
the basis of heuristic arguments,~\cite{diep-johnson00} and later
generalized by Schenter to the case of rigid bodies and applied to a
model for water.~\cite{Schenter02}

Equation~(\ref{eq:U2_eff}) is actually the discretized version of a
path integral, as shown in Eq.~(\ref{eq:U2_pi}). 
The circled integral is defined as
\begin{eqnarray}
  \oint {\cal D} \boldsymbol x ~ \exp\left[ -\frac{1}{\hbar} \int_0^{\beta \hbar}
  \frac{m}{2} 
  \left| \frac{\mathrm d \boldsymbol x(\tau)}{\mathrm d \tau} \right|^2 ~ 
  \mathrm d \tau \right] &\equiv& \nonumber \\
  \lim_{P \rightarrow \infty}
  \int \prod_{i=1}^{P-1} \mathrm d\Delta\boldsymbol{x}^{(i)} 
  F_\mathrm{ring}(\Delta\boldsymbol{x}^{(1)},\ldots,\Delta\boldsymbol{x}^{(P)})
  &=& 1,
\label{eq:pi}
\end{eqnarray}
and it indicates that one has to consider all the cyclic paths with
ending points at the origin, that is $\boldsymbol x(0) = \boldsymbol
x(\beta \hbar) = \boldsymbol 0$. The normalization of the path
integral is also indicated in Eq.~(\ref{eq:pi}).

We can perform on Eq.~(\ref{eq:Bx}), describing the exchange
contribution to the second virial coefficient, the same steps leading
from Eq.~(\ref{eq:Bdir}) to Eq.~(\ref{eq:BBoltz}). The only difference
is the presence of the permutation operator, whose main consequence is
fact that $\boldsymbol r^{(P+1)}_1 = \boldsymbol r^{(1)}_2$ and
$\boldsymbol r^{(P+1)}_2 = \boldsymbol r^{(1)}_1$. In this case,
defining $\boldsymbol X^{(i)} = \boldsymbol r^{(i)}_1$ and
$\boldsymbol X^{(i+P)} = \boldsymbol r^{(i)}_2$, one obtains
\begin{widetext}
\begin{eqnarray}
  B_\mathrm{xc}(T) &=& -\frac{(-1)^{2I} \Lambda^6}{2 (2I+1) V} \int
  \mathrm d \boldsymbol X^{(1)} \ldots \mathrm d \boldsymbol X^{(2P)} ~
  \exp\left[-\frac{\beta}{P} 
    \sum_{i=1}^P U_2(|\boldsymbol X^{(P+i)} - \boldsymbol X^{(i)}|) \right] 
\times \nonumber \\ & & 
\left(
  \frac{P^{3/2}}{\Lambda^3} \right)^{2P} \exp\left[-\frac{\pi
      P}{\Lambda^2} \sum_{i=1}^{2P} \left( \boldsymbol X^{(i+1)} - \boldsymbol
    X^{(i)}\right)^2 \right]
\label{eq:Bxc} \\
&=&
- \frac{(-1)^{2I} \Lambda^3}{2 (2I+1) V} \int \mathrm d \boldsymbol X^{(1)} \ldots \mathrm d \boldsymbol X^{(2P)} ~ 
\exp\left[-\frac{\beta}{P} 
\sum_{i=1}^P U_2(|\boldsymbol X^{(P+i)} - \boldsymbol X^{(i)}|)  \right] 
\times \nonumber \\ & &
\frac{\Lambda_\mu^3}{2^{3/2}}  \left( \frac{(2P)^{3/2}}{\Lambda_\mu^3}
  \right)^{2P} 
\exp\left[-\frac{\pi ~ 2 P}{\Lambda_\mu^2} \sum_{i=1}^{2P} 
  \left( \boldsymbol X^{(i+1)} - \boldsymbol X^{(i)}\right)^2 \right] \\
&=& 
-\frac{1}{2} \frac{(-1)^{2I}\Lambda^3}{(2I+1) ~ 2^{3/2}} \left\langle
\exp\left[-\frac{\beta}{P} 
 \sum_{i=1}^P U_2(|\boldsymbol X^{(P+i)} - \boldsymbol X^{(i)}|)  \right]
\right\rangle 
\label{eq:Bexchange} \\
&\underset{P \rightarrow \infty}{=}&
- \frac{(-1)^{2I}\Lambda^3}{(2I+1)~2^{5/2}} ~ \oint {\cal D}
   \boldsymbol X ~ \exp\left[
     -\frac{1}{\hbar} \int_0^{\beta \hbar} \frac{\mu}{2} \left|
     \frac{\mathrm d \boldsymbol X(\tau)}{\mathrm d \tau} \right|^2 +
     U_2(|\boldsymbol X(\tau + \beta \hbar /2) - \boldsymbol X(\tau)|)
     ~ \mathrm d\tau
 \right],
 \label{eq:Bexch_pi}
\end{eqnarray}  
\end{widetext}
where we have defined $\Lambda_\mu =
\sqrt{2}\Lambda$. The exchange contribution to the second virial
coefficient is given simply as an average of the two-body potential
taken on ring polymers corresponding to particles of mass $\mu = m/2$.
In the discretized version of the path integral, one has to consider
$2P$ beads.
In Eq.~(\ref{eq:Bexchange}), we have used the overall translation
invariance of the integral to remove the factor of $V$ in the
denominator.

The effect of the various permutations can be visualized as generating paths
with a larger number of beads, which are obtained by coalescing the
ring polymers corresponding to the particles that are exchanged by the
permutation operator.

\subsection{Third virial coefficient}

We now discuss the third virial coefficient, starting from the
expression given in Eq.~(\ref{eq:C}). Since $4 B^2(T)$ can be
calculated by the methods of the previous section, we concentrate on
the second term, whose summands can be written as follows:
\begin{eqnarray}
Z_3     &=& \Lambda^9 \int \mathrm d1 \mathrm d2 \mathrm d3 ~ 
\langle 1 2 3 | \mathrm e^{-\beta \hat H_3} \sum_{\pi_3} {\cal P}_{\pi_3} | 1 2 3
\rangle \\ 
Z_2 Z_1 &=& \Lambda^9 \int \mathrm d1 \mathrm d2 \mathrm d3 ~
\langle 1 2 3 | \mathrm e^{-\beta (\hat H_2 + \hat T_3)} \sum_{\pi_2} {\cal P}_{\pi_2}
| 1 2 3 \rangle \label{eq:Z2Z1} \\ 
Z_1^3   &=& \Lambda^9 \int \mathrm d1 \mathrm d2 \mathrm d3 ~
\langle 1 2 3 | \mathrm e^{-\beta \hat K_3} | 1 2 3 \rangle,  
\label{eq:Z13}
\end{eqnarray}
where $\hat T_i = -\frac{\hbar^2}{2m} \nabla_i^2$ is the kinetic energy
operator of particle $i$.

We can simplify the expression in square brackets on the right-hand side of
Eq.~(\ref{eq:C}) by writing the three $Z_2 Z_1$ terms choosing each time a
different particle for $Z_1$ (in Eq.~(\ref{eq:Z2Z1}) we have chosen particle 3
as coming from $Z_1$).  After considering all the permutations of two and
three particles, we end up with $6 + 3 \times 2 + 1 = 13$ terms
building the term in square brackets of Eq.~(\ref{eq:C}). It is useful to
collect these 13 terms as follows:
\begin{enumerate}
\item{{\em Term 1 (identity term)}: we sum together permutation $123$ from $Z_3$, the
  identity permutations from the three $Z_2 Z_1$ and  the whole $2 Z_1^3$ term.
  Adding $4 B^2_{\rm Boltzmann}(T)$, one obtains the
  Boltzmann expression for $C(T)$, already discussed in
  Ref.~\onlinecite{Garberoglio2009b}. In the present formulation based on Cartesian
  coordinates, the value $C(T)$ in the case of Boltzmann statistics involves
  an average over three independent ring polymers, which correspond to the
  three particles. In the following, this contribution to $C(T)$ will be
  referred to as $C_{\rm Boltzmann}(T)$ and is made by $1 + 3 + 1 = 5$ of the
  13 terms described above.}
\item{{\em Term 2 (odd term)}: we take permutations $132$, $213$ and $321$
  from $Z_3$ and the three exchange permutations from the $Z_2 Z_1$
  terms. These permutations are all odd, and they have to be multiplied by
  $(-1)^{2I}/(2I+1)$.  All of these
  permutations correspond to configurations where two of the three particles
  are exchanged.  The sum of these 6 terms will be referred to as $C_{\rm
    odd}(T)$.}
\item{{\em Term 3 (even term)}: we take the permutations $231$ and $312$ from
  $Z_3$. These are the remaining two terms from the 13, and
  are both even permutations, hence the name. Both of these terms
  correspond to a cyclic exchange of the three particles, and their
  sum will be referred to as $C_{\rm even}(T)$. They are to be weighted
  with $1/(2I+1)^2$.}
\end{enumerate}

Using these definitions, the full $C(T)$, including quantum statistical
effects, can be written as
\begin{equation}
  C(T) = C_{\rm Boltzmann}(T) +
  (-1)^{2I} \frac{C_{\rm odd}(T)}{2I+1}
  + \frac{C_{\rm even}(T)}{(2I+1)^2} +
  C_{\rm B}(T),
\label{eq:C_full}
\end{equation}
where the last term in the right-hand sum is given by
\begin{equation}
  C_{\rm B}(T) = 8 B_{\rm Boltzmann}(T) B_\mathrm{xc}(T) + 4  B^2_\mathrm{xc}(T),
\label{eq:CB}
\end{equation}
since the contribution of $4 B^2_\mathrm{Boltzmann}(T)$ to $C(T)$ is
already included in $C_{\rm Boltzmann}(T)$. In Eqs.~(\ref{eq:C_full})
and (\ref{eq:CB}), the upper (lower) sign corresponds to Bose--Einstein
(Fermi--Dirac) statistics.

Using the same procedure outlined above in the case of $B(T)$, one can
write the Boltzmann contribution to the third virial coefficient as
\begin{widetext}
\begin{eqnarray}
  C_\mathrm{Boltzmann}(T) &=& 4 B^2_\mathrm{Boltzmann}(T) - \nonumber
  \\
& & \frac{1}{3} 
  \int \mathrm d\boldsymbol{r}_1 \mathrm d\boldsymbol{r}_2
\left[
\mathrm e^{-\beta \overline{V}_3(\boldsymbol{r}_1,\boldsymbol{r}_2)} -
\mathrm e^{-\beta \overline{U}_2(|\boldsymbol{r}_1|)} -
\mathrm e^{-\beta \overline{U}_2(|\boldsymbol{r}_2|)} -
\mathrm e^{-\beta \overline{U}_2(|\boldsymbol{r}_1 - \boldsymbol{r}_2|)}
+ 2 \right],
\label{eq:CBoltz} \\
\exp\left[-\beta \overline{V}_3(\boldsymbol{r}_1,\boldsymbol{r}_2)\right]
&=&
\int \prod_{i=1}^{P-1}
\Delta \boldsymbol x_1^{(i)} 
\Delta \boldsymbol x_2^{(i)} 
\Delta \boldsymbol x_3^{(i)} ~
F_\mathrm{ring}^{(1)}
F_\mathrm{ring}^{(2)}
F_\mathrm{ring}^{(3)} ~
\exp\left[-\beta 
\overline{V}_3^{\mathrm B}(\boldsymbol{r}_1,\boldsymbol{r}_2)
\right]
\label{eq:V3disc}
\\
&\underset{P \rightarrow \infty}{=}&
\oint {\cal D} \boldsymbol x_1
      {\cal D} \boldsymbol x_2 
      {\cal D} \boldsymbol x_3 ~ 
\exp\left[-\frac{1}{\hbar} \int_0^{\beta \hbar} 
\frac{m}{2} \left(
\left| \frac{\mathrm d \boldsymbol x_1(\tau)}{\mathrm d \tau} \right|^2 +
\left| \frac{\mathrm d \boldsymbol x_2(\tau)}{\mathrm d \tau} \right|^2 +
\left| \frac{\mathrm d \boldsymbol x_3(\tau)}{\mathrm d \tau} \right|^2 
\right) + \right. 
\nonumber \\
& & \left.
V_3(\boldsymbol r_1 + \boldsymbol x_1(\tau),
\boldsymbol r_2 + \boldsymbol x_2(\tau), \boldsymbol x_3(\tau))
~ \mathrm d \tau
\right], \label{eq:V3}
\end{eqnarray}  
\end{widetext}
where $F_\mathrm{ring}^{(k)}$ denotes the probability distribution of
the path relative to particle $k$, as defined in Eq.~(\ref{eq:Fring}).
In Eq.~(\ref{eq:V3disc}), the three-body effective potential energy
$\overline{V}_3$ is obtained as an average performed over {\em three}
independent ring polymers of the total three-body potential energy:
\begin{eqnarray}
V_3(\boldsymbol{x}, \boldsymbol{y}, \boldsymbol{z}) &=& 
U_3(\boldsymbol{x}, \boldsymbol{y}, \boldsymbol{z}) + 
U_2(|\boldsymbol x - \boldsymbol y|) +
\nonumber \\
& & 
U_2(|\boldsymbol x - \boldsymbol z|) +
U_2(|\boldsymbol y - \boldsymbol z|),
\end{eqnarray}
where $U_3(\boldsymbol{x}, \boldsymbol{y}, \boldsymbol{z})$ is the
non-additive three-body potential of three atoms.
In Eq.~(\ref{eq:V3disc}) the total three-body
potential energy for the Boltzmann contribution to the third virial
coefficient is
\begin{eqnarray}
\overline{V}_3^{\mathrm B}(\boldsymbol{r}_1,\boldsymbol{r}_2) &=&
\frac{1}{P} \sum_{i=1}^P U_3(\boldsymbol{r}_1 + \boldsymbol{x}_1^{(i)},
\boldsymbol{r}_2 + \boldsymbol{x}_2^{(i)},
\boldsymbol{x}_3^{(i)}) + \nonumber \nonumber \\ & &
U_2(|\boldsymbol{r}_1 + \boldsymbol{x}_1^{(i)} - 
     \boldsymbol{r}_2 - \boldsymbol{x}_2^{(i)}|) +
\nonumber \nonumber \\ & &
U_2(|\boldsymbol{r}_1 + \boldsymbol{x}_1^{(i)} - 
     \boldsymbol{x}_3^{(i)}|) +\nonumber \nonumber \\ & &
U_2(|\boldsymbol{r}_2 + \boldsymbol{x}_2^{(i)} - \boldsymbol{x}_3^{(i)}|),
\end{eqnarray}
where the variables with superscript $(i)$ denote the coordinates of
three ring polymers with one of the beads at the origin.  Notice that
in passing from Eq.~(\ref{eq:C}) to Eq.~(\ref{eq:CBoltz}) we have used
the translation invariance of the integrand to perform the integration
over $\boldsymbol r_3$, which removed the factor of $V$ in the
denominator. As a consequence, the paths corresponding to particle 3
have their endpoints at the origin of the coordinate system (or,
equivalently, the third particle is fixed at the origin when the
classical limit is performed.)  In the same limit, the variables
$\boldsymbol r_1$ and $\boldsymbol r_2$ appearing in Eq.~(\ref{eq:V3})
reduce to the positions of particles 1 and 2, respectively, and one has
$\overline V_3(\boldsymbol r_1,\boldsymbol r_2) = V_3(\boldsymbol
r_1,\boldsymbol r_2)$.

The term $C_\mathrm{odd}(T)$ is obtained by exchanging the positions
of two particles. This operation reduces the number of ring polymers
to two: one having $2P$ beads, corresponding to the exchanged
particles, and the other having $P$ beads, corresponding to the
remaining one. The odd contribution is given by
\begin{widetext}
\begin{eqnarray}
  C_\mathrm{odd}(T) &=& - \frac{\Lambda^9}V 
\int \mathrm d1 \mathrm d2 \mathrm d3 ~ 
\langle 1 2 3 | 
\exp\left[-\beta \hat H_3 \right] - \exp\left[-\beta (\hat
  K_2 + \hat U_2(\boldsymbol{r}_2 - \boldsymbol{r}_1)) \right]
| 2 1 3 \rangle  \label{eq:Codd1} \\
&=&
- \frac{\Lambda^3}{2^{3/2}} \int \mathrm d \boldsymbol r_3 ~ \left\langle
\exp\left[ -\beta \overline{V}^\mathrm{odd}_3 \right] - \exp\left[ -\beta
  \overline{U}^\mathrm{odd}_2 \right]
\right\rangle
\label{eq:Codd}  \\
&=&
- \frac{\Lambda^3}{2^{3/2}} \int \mathrm d \boldsymbol r_3 ~
\left\{
\oint {\cal D}\boldsymbol x {\cal D}\boldsymbol y ~ 
\exp\left[-\frac{1}{\hbar}
\int_{0}^{\beta \hbar}
\frac{m}{4} \left| \frac{\mathrm d \boldsymbol x(\tau)}{\mathrm d
  \tau} \right|^2 +
\frac{m}{2} \left| \frac{\mathrm d \boldsymbol y(\tau)}{\mathrm d
  \tau} \right|^2
+ V_3\left(\boldsymbol x\left(\tau + {\beta \hbar}/{2}\right),
      \boldsymbol x(\tau), \boldsymbol r_3 + \boldsymbol y(\tau)\right)
\mathrm d \tau \right] \right. \nonumber \\
& & \left.
- \oint {\cal D}\boldsymbol x ~ \exp\left[-\frac{1}{\hbar}
\int_{0}^{\beta \hbar}
\frac{m}{4} \left| \frac{\mathrm d \boldsymbol x(\tau)}{\mathrm d
  \tau} \right|^2
+ U_2(|\boldsymbol x(\tau + \beta \hbar /2) - \boldsymbol x(\tau)|)
\mathrm d \tau \right]
\right\},
\end{eqnarray}
\end{widetext}
where we have defined
\begin{eqnarray}
  \overline{V}^\mathrm{odd}_3(\boldsymbol r_3) &=& \frac{1}{P} \sum_{i=1}^{P} 
  U_3(\boldsymbol X^{(i)}, \boldsymbol X^{(i+P)}, 
  \boldsymbol r_3 + \boldsymbol x_3^{(i)}) + 
\nonumber \\ & &
  U_2(|\boldsymbol X^{(i)} - \boldsymbol X^{(i+P)}|) + 
\nonumber \\ & &
  U_2(|\boldsymbol X^{(i)} - \boldsymbol r_3 - \boldsymbol x_3^{(i)}|) + 
\nonumber \\ & &
  U_2(|\boldsymbol X^{(i+P)} - \boldsymbol r_3 - \boldsymbol x_3^{(i)}|) 
\label{eq:V_odd} \\
\overline{U}^\mathrm{odd}_2 &=& 
\frac{1}{P} \sum_{i=1}^{P} 
    U_2(|\boldsymbol X^{(i)} - \boldsymbol X^{(i+P)}|).
\label{eq:U2_odd}
\end{eqnarray}
The $2P$ variables $\boldsymbol X^{(i)}$ have been defined analogously
to what has been done in Eq.~(\ref{eq:Bxc}).  Notice that in the
discretized version, the average defining the odd exchange term in
Eq.~(\ref{eq:Codd}) is performed over two different kinds of ring
polymers: the first has $2P$ beads of mass $m/2$ and connects
particles 1 and 2 whose coordinates are exchanged by the permutation
operator, whereas the second -- corresponding to the third particle of
mass $m$ -- has $P$ beads.

A similar derivation holds for the even contribution to the third virial
coefficient, which is given by
\begin{widetext}
\begin{eqnarray}
  C_\mathrm{even}(T) &=& -\frac{2 \Lambda^9}{3 V}  
\int \mathrm d1 \mathrm d2 \mathrm d3 ~ 
\left\langle 1 2 3 \left|
  \exp\left( -\beta \hat H_3 \right) \right| 3 1 2 \right\rangle 
= 
-\frac{2}{3}  \frac{\Lambda^6}{3^{3/2}} \left\langle
\exp\left( - \beta \overline{V}^\mathrm{even}_3 \right)
\right\rangle
\label{eq:Ceven} \\
 &=&
-\frac{2}{3}  \frac{\Lambda^6}{3^{3/2}}
\oint {\cal D}\boldsymbol{x} ~ \exp\left[
-\frac{1}{\hbar} \int_0^{\beta \hbar} \frac{m}{6} 
\left| \frac{\mathrm d \boldsymbol x(\tau)}{\mathrm d \tau} \right|^2
+ V_3(\boldsymbol x(\tau + 2 \beta \hbar /3),
      \boldsymbol x(\tau + \beta \hbar /3),
      \boldsymbol x(\tau))
\mathrm d \tau \right],
\end{eqnarray}
\end{widetext}
where we have defined
\begin{eqnarray}
  \overline{V}^\mathrm{even}_3 &=& \frac{1}{P} \sum_{i=1}^{P} 
  U_3(\boldsymbol Y^{(i)}, \boldsymbol Y^{(i+P)}, \boldsymbol Y^{(i+2P)}) + 
\nonumber \\ & &
  U_2(|\boldsymbol Y^{(i)} - \boldsymbol Y^{(i+P)}|) + 
\nonumber \\ & &
  U_2(|\boldsymbol Y^{(i)} - \boldsymbol Y^{(i+2P)}|) + 
\nonumber \\ & &
  U_2(|\boldsymbol Y^{(i+P)} - \boldsymbol Y^{(i+2P)}|),
\label{eq:V_even}
\end{eqnarray}
together with
$\boldsymbol Y^{(i)} = \boldsymbol r_1^{(i)}$, 
$\boldsymbol Y^{(i+P)} = \boldsymbol r_2^{(i)}$, and
$\boldsymbol Y^{(i+2P)} = \boldsymbol r_3^{(i)}$.
In the discretized version, the even contribution to the third virial
coefficient is an average over the coordinates of the $3P$ beads of a
single ring polymer corresponding to a particle of mass $m/3$.

Notice that, from a computational point of view, the evaluation of the
exchange contributions to the third virial coefficient is much less
demanding than the calculation of the Boltzmann part, which is
given as an integral over the positions of two particles. In fact, the
odd contribution is calculated as an integration over the position of
one particle only, whereas the even contribution is given by a simple
average over ideal-gas ring-polymer configurations.  In particular,
the full calculation of $C(T)$ at the lowest temperature with 2.5~GHz
processors required $\sim 2400$ CPU hours, only 15\% of which was
needed to calculate the exchange contributions.

\section{Results and discussion}

\subsection{Details of the calculation}

We have calculated $C(T)$ for both isotopes of helium with the
path-integral method described above. We used the highly accurate
two-body potential of Przybytek {\em et al.},~\cite{He2-2010} which
includes the most significant corrections (adiabatic, relativistic,
and quantum electrodynamics) to the Born--Oppenheimer result. We also
used the three-body {\em ab initio} potential of Cencek {\em et
  al.},~\cite{CPS09} which was derived at the Full Configuration
Interaction level and has an uncertainty approximately one-fifth that of
the three-body potential~\cite{He-3body} used in our
previous work.~\cite{Garberoglio2009b}

We generated ring-polymer configurations using the interpolation
formula of Levy.~\cite{levy54,fosdick-jordan66}
The number of beads was chosen as a function of the temperature $T$
according to the formulae $P = \mathrm{int}[(1200~\mathrm{K})/T] + 7$ for ${}^4$He and $P = \mathrm{int}[(1800~\mathrm{K})/T] + 7$ for ${}^3$He, where $\mathrm{int}[x]$ indicates the integer closest to $x$. These values of $P$ were
enough to reach convergence in the path-integral results at all the temperatures considered in the present study.
The spatial integrations were performed with the VEGAS
algorithm~\cite{vegas1}, as implemented in the GNU Scientific
Library,~\cite{gsl} with 1 million integration points and cutting off
the interactions at 4~nm.
The three-body interaction was pre-calculated on a three-dimensional
grid and interpolated with cubic splines.
The values of the virial coefficient and their statistical uncertainty
were obtained by averaging over the results of 256 independent
runs.

First of all, we checked that our methodology was able to reproduce
well-converged fully quantum $B(T)$ calculations for helium, which
were obtained using the same pair potential as the present
work.~\cite{Cencek11} Our results agree within mutual uncertainties
with these independent calculations, and confirm the observation,
already made when analyzing theoretical $B(T)$ calculations performed
using Lennard-Jones potentials, that exchange effects are significant
only for temperatures lower than about 7~K.~\cite{boyd69} The exchange
contribution to the second virial coefficient is negative in the case
of Bose--Einstein statistics and positive in the case of Fermi--Dirac
statistics, as one would expect.

\subsection{The third virial coefficient of ${}^4$He}

\begin{table*}
  \begin{center}
  \begin{tabular}{d|dd|dd|dd|dd|dd}
\multicolumn{1}{c|}{Temperature} & 
\multicolumn{2}{c|}{$C$} & 
\multicolumn{2}{c|}{$C_\mathrm{Boltzmann}$} & 
\multicolumn{2}{c|}{$C_\mathrm{odd}$} &
\multicolumn{2}{c|}{$C_\mathrm{even}$} & 
\multicolumn{2}{c}{$C_\mathrm{B}$} \\
(\mathrm K) & 
\multicolumn{2}{c|}{$(\mathrm{cm}^6~\mathrm{mol}^{-2})$} &
\multicolumn{2}{c|}{$(\mathrm{cm}^6~\mathrm{mol}^{-2})$} &
\multicolumn{2}{c|}{$(\mathrm{cm}^6~\mathrm{mol}^{-2})$} & 
\multicolumn{2}{c|}{$(\mathrm{cm}^6~\mathrm{mol}^{-2})$} & 
\multicolumn{2}{c}{$(\mathrm{cm}^6~\mathrm{mol}^{-2})$} \\
\hline 
2.6 & 266 & \pm 21 & 245 & \pm 21 &  -863 & \pm 3 & -88.1 & \pm
0.5 & 972 & \pm 1 \\
2.8 & 631 & \pm 21  & 607 & \pm 20 & -504 & \pm 2 & -49.0 & \pm 0.3 &
577.5 & \pm 0.7 \\
3 & 848 & \pm 17 & 828 & \pm 17 & -301.7 & \pm 1.4 & -28.0 & \pm 0.2 &
 349.67 & \pm 0.5 \\
3.2 & 937 & \pm 14 & 923 & \pm 14 & -184.9 & \pm 0.8 & -16.29 & \pm 0.12 &
215.0 & \pm 0.3 \\
3.5 & 1061 & \pm 10 & 1050 & \pm 10 & -88.9 & \pm 0.5 & -7.35 & \pm 0.06 &
106.89 & \pm 0.15 \\
3.7 & 1070 & \pm 9 & 1062 & \pm 9 & -55.5 & \pm 0.4 & -4.50 & \pm 0.05 &
67.97 & \pm 0.12 \\
4 & 1082 & \pm 8  & 1077 & \pm 8 & -27.9 & \pm 0.2 & -2.14 & \pm 0.02
& 35.14 & \pm 0.07 \\
4.2 & 1074 & \pm 7 & 1070 & \pm 7 & -17.79 & \pm 0.16 &
-1.352 & \pm 0.017 & 22.83 & \pm 0.05 \\
4.5 & 1049 & \pm 6 & 1047 & \pm 6 & -9.156 & \pm 0.09 & 
-0.663 & \pm 0.008 & 12.18 & \pm 0.03 \\
5 & 986 & \pm 5 & 985 & \pm 5 & -3.28 & \pm 0.05 & -0.227 & \pm 0.004 
& 4.50 & \pm 0.02 \\
6 & 861 & \pm 3 & 861 & \pm 3 & -0.361 & \pm 0.014 & -0.027 & \pm
0.001 & 0.682 & \pm 0.004 \\
7 & 746& \pm 2 & 746 & \pm 2 & 
-0.029 & \pm 0.003 &
-0.004 & \pm 0.0003 &
0.115  & \pm 0.001 \\
8.5     & 620.7 & \pm 1.6 & 620.7 & \pm 1.6 & & & & & & \\
10      & 532.3 & \pm 0.8 & 532.3 & \pm 0.8 & & & & & & \\
12      & 449.7 & \pm 0.8 & 449.7 & \pm 0.8 & & & & & & \\
13.8033 & 401.0 & \pm 0.4 & 401.0 & \pm 0.4 & & & & & & \\
15      & 375.1 & \pm 0.5 & 375.1 & \pm 0.5 & & & & & & \\
17      & 342.2 & \pm 0.4 & 342.2 & \pm 0.4 & & & & & & \\
18.689  & 321.2 & \pm 0.2 & 321.2 & \pm 0.2 & & & & & & \\
20      & 307.7 & \pm 0.3 & 307.7 & \pm 0.3 & & & & & & \\
24.5561 & 274.2 & \pm 0.2 & 274.2 & \pm 0.2 & & & & & & 
  \end{tabular}
\caption{Values of the third virial coefficient of ${}^4$He and its
  components at selected temperatures. 
The $\pm$ values reflect only the standard uncertainty of the Monte
Carlo integration; see Ref.~\onlinecite{paper2} for complete
uncertainty analysis.}
\label{tab:CHe4}      
\end{center}
\end{table*}  

\begin{figure}
\includegraphics[width=0.95\linewidth]{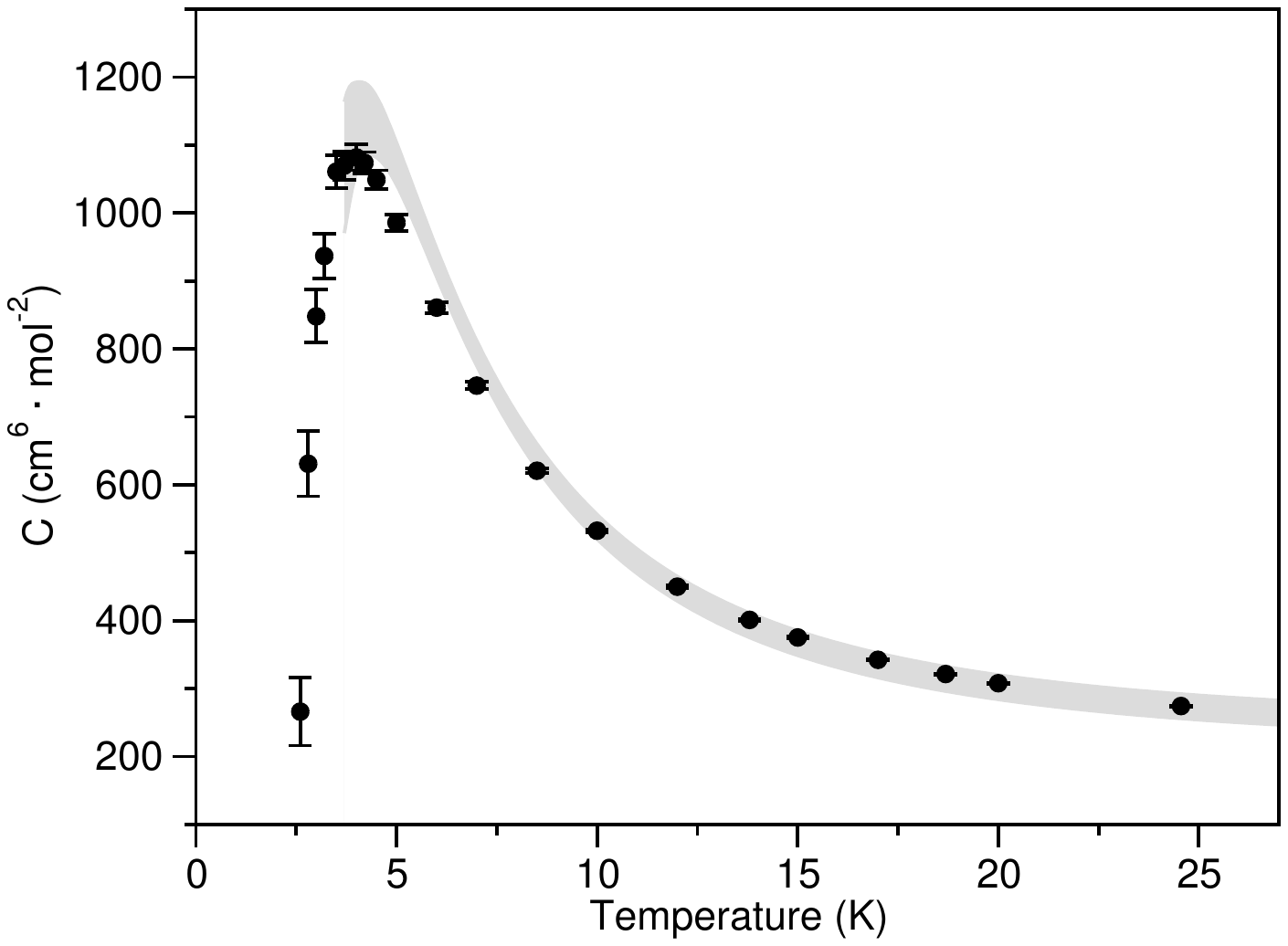}
\caption{The third virial coefficient of ${}^4$He. The black circles
  are the results of the present calculations, with error bars
  representing expanded uncertainties with coverage factor $k=2$. The
  gray area shows the results of the recent low-temperature
  experiments by Gaiser and collaborators.~\cite{Gaiser09,Gaiser10}}
\label{fig:CHe4}  
\end{figure}

We report in Table~\ref{tab:CHe4} the values of the third virial
coefficient of ${}^4$He, together with the various contributions of
Eq.~(\ref{eq:C_full}), for temperatures in the range from $2.6$~K to
$24.5561$~K, which is the lowest temperature studied in our
previous work.~\cite{Garberoglio2009b}
The same data are plotted in Figure~\ref{fig:CHe4}, where they are
compared with the recent experimental measurements by Gaiser and
collaborators.~\cite{Gaiser09,Gaiser10}

More extensive comparison with available data over a wide range of
temperatures will be presented elsewhere.~\cite{paper2}
In Fig.~\ref{fig:CHe4}, our results are plotted with expanded
uncertainties with coverage factor $k=2$ as derived in
Ref.~\onlinecite{paper2}; the uncertainty at the same expanded level
for the experimental results was estimated from a figure in
Ref.~\onlinecite{Gaiser09}.

First, we notice that exchange effects are completely
negligible in the calculation of the third virial coefficient for
temperatures larger than 7~K, where their contribution to the overall
value is close to one thousandth of that of the Boltzmann part.
This is analogous to what has already been observed for the second virial
coefficient.

When the temperature is lower than 7~K, the various exchange terms
have contributions of similar magnitude and opposite sign, but their
overall contribution to $C(T)$ is positive at all the temperatures 
that have been investigated. The exchange contribution to $C(T)$
is comparable to the statistical uncertainty of the calculation, which
progressively increases as the temperature is lowered.

In Fig.~\ref{fig:CHe4}, it can be seen that our theoretical values of
$C(T)$ are compatible with those of recent
experiments~\cite{Gaiser09,Gaiser10} down to the temperature of
10~K. For lower temperatures, the experimental results are
somewhat larger than the calculated values, even though agreement
is found again for temperatures below 4~K, where $C(T)$ passes through
a maximum.

\subsection{The third virial coefficient of ${}^3$He}

Similar behavior is observed in the case of the third virial
coefficient for ${}^3$He, whose calculated values are reported in
Table~\ref{tab:CHe3}. Also in this case the exchange contributions are
of opposite signs, but their combined effect is to reduce the value
obtained with Boltzmann statistics, which is the opposite trend to
that observed for ${}^4$He.

\begin{figure}
\includegraphics[width=0.95\linewidth]{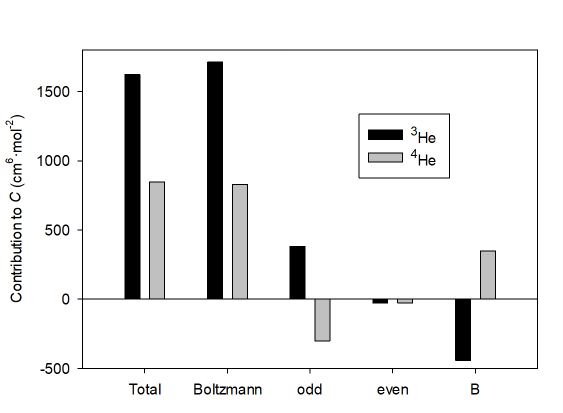}
\caption{The magnitude and sign of the various contributions to $C(T)$
at $T=3$~K.}
\label{fig:BarChart}  
\end{figure}

The effects of the various contributions to the third virial
coefficient, in both the Bose--Einstein and Fermi--Dirac case, are
summarized in Fig.~\ref{fig:BarChart} for the representative
temperature of $T=3$~K.  First, we notice that the largest
contribution to the third virial coefficient comes from the Boltzmann
term. The even exchange term has only a minor contribution, whereas
the two remaining terms ($C_\mathrm{odd}$ and $C_\mathrm B$)
have almost equal magnitudes and opposite signs.  In the case of
Bose--Einstein statistics, the contribution to $C$ from
$C_\mathrm{odd}$ is negative, while that from $C_\mathrm B$ is
positive; the opposite situation is observed in the case of
Fermi--Dirac statistics.  The overall sum of the exchange
contributions is positive for ${}^4$He and negative in the case of
${}^3$He.

The magnitude of each exchange contribution at a given temperature is
significantly greater for ${}^3$He; this reflects the larger de~Broglie
wavelength, which not only appears directly in the exchange terms but
also affects the range of space sampled by the ring polymers.

In the case of ${}^3$He, the exchange contribution is significantly
larger than the uncertainty of our calculations, at least at the
lowest temperatures that we have investigated. Similarly to the case
of ${}^4$He, quantum statistical effects on $C(T)$ contribute less
than one part in a thousand for temperatures higher than 7~K.
Even in the case of ${}^3$He, we observe $C(T)$ pass through a
maximum, at a temperature around 3~K, which is 1~K lower than the
temperature where $C(T)$ reaches a maximum for the ${}^4$He isotope.

\begin{table*}
  \begin{center}
  \begin{tabular}{d|dd|dd|dd|dd|dd}
\multicolumn{1}{c|}{Temperature} & 
\multicolumn{2}{c|}{$C$} & 
\multicolumn{2}{c|}{$C_\mathrm{Boltzmann}$} & 
\multicolumn{2}{c|}{$C_\mathrm{odd}$} &
\multicolumn{2}{c|}{$C_\mathrm{even}$} & 
\multicolumn{2}{c}{$C_\mathrm{B}$} \\
(\mathrm K) & 
\multicolumn{2}{c|}{$(\mathrm{cm}^6~\mathrm{mol}^{-2})$} &
\multicolumn{2}{c|}{$(\mathrm{cm}^6~\mathrm{mol}^{-2})$} &
\multicolumn{2}{c|}{$(\mathrm{cm}^6~\mathrm{mol}^{-2})$} & 
\multicolumn{2}{c|}{$(\mathrm{cm}^6~\mathrm{mol}^{-2})$} & 
\multicolumn{2}{c}{$(\mathrm{cm}^6~\mathrm{mol}^{-2})$} \\
\hline 
2.6	&	1657	&	\pm	29	&	1857	&	\pm	28	&	-1803	&	\pm	4	&	-274.8	&	\pm	0.8	&	-1033	&	\pm	5	\\
2.8	&	1686	&	\pm	23	&	1817	&	\pm	23	&	-1164	&	\pm	3	&	-167.9	&	\pm	0.6	&	-671	&	\pm	4	\\
3	&	1622	&	\pm	17	&	1712	&	\pm	17	&	-760.2	&	\pm	2.2	&	-105.7	&	\pm	0.4	&	-443.3	&	\pm	2.2	\\
3.2	&	1561	&	\pm	17	&	1621	&	\pm	17	&	-503.5	&	\pm	1.7	&	-66.8	&	\pm	0.3	&	-295	&	\pm	1.6	\\
3.5	&	1451	&	\pm	13	&	1487	&	\pm	13	&	-277.2	&	\pm	1.2	&	-34.89	&	\pm	0.15	&	-165.8	&	\pm	1.1	\\
3.7	&	1412	&	\pm	11	&	1439	&	\pm	11	&	-189.2	&	\pm	0.8	&	-23.12	&	\pm	0.11	&	-115.5	&	\pm	1	\\
4	&	1326	&	\pm	9	&	1342	&	\pm	9	&	-107	&	\pm	0.5	&	-12.69	&	\pm	0.07	&	-66.5	&	\pm	0.7	\\
4.2	&	1261	&	\pm	9	&	1273	&	\pm	9	&	-73.9	&	\pm	0.5	&	-8.65	&	\pm	0.05	&	-46.8	&	\pm	0.2	\\
4.5	&	1183	&	\pm	7	&	1190	&	\pm	7	&	-43.7	&	\pm	0.3	&	-4.86	&	\pm	0.03	&	-27.9	&	\pm	0.2	\\
5	&	1075	&	\pm	6	&	1079	&	\pm	6	&	-18.41	&	\pm	0.16	&	-1.963	&	\pm	0.016	&	-12.31	&	\pm	0.13	\\
6	&	896	&	\pm	4	&	897	&	\pm	4	&	-3.56	&	\pm	0.06	&	-0.353	&	\pm	0.005	&	-2.52	&	\pm	0.04	\\
7	&	773	&	\pm	3	&	773	&	\pm	3	&	-0.78	&	\pm	0.02	&	-0.0784	&	\pm	0.002	&	-0.6	&	\pm	0.01	\\
8.5	&	645	&	\pm	2	&	645	&	\pm	2	&	-0.059	&	\pm	0.006	&	-0.0087	&	\pm	0.0003	&	-0.072	&	\pm	0.003	\\
10 & 558.3 &\pm  1.6 & 558.3 &\pm  1.6 & & & & & & \\
12 & 475.5 &\pm  1.1 & 475.5 &\pm  1.1 & & & & & & \\
13.8033 & 426.2   &\pm  0.8    & 426.2   &\pm  0.8    & & & & & & \\
15 & 402.0 &\pm  0.7 & 402.0 &\pm  0.7 & & & & & & \\
17 & 369.6 &\pm  0.5 & 369.6 &\pm  0.5 & & & & & & \\
18.689  & 347.8   &\pm  0.4    & 347.8   &\pm  0.4    & & & & & & \\
20      & 333.4   &\pm  0.4    & 333.4   &\pm  0.4    & & & & & & \\
24.5561 & 297.8   &\pm  0.3    & 297.8   &\pm  0.3    & & & & & & 
\end{tabular}
\caption{Values of the third virial coefficient of ${}^3$He and its
  components at selected temperatures. Note that the various contributions
  should be summed with the weights appearing in Eq.~(\ref{eq:C_full}) with
  $I=1/2$.  The $\pm$ values reflect only the standard
  uncertainty of the Monte Carlo integration; see
  Ref.~\onlinecite{paper2} for complete uncertainty analysis.}
\label{tab:CHe3}      
\end{center}
\end{table*}  

\begin{figure}
\includegraphics[width=0.95\linewidth]{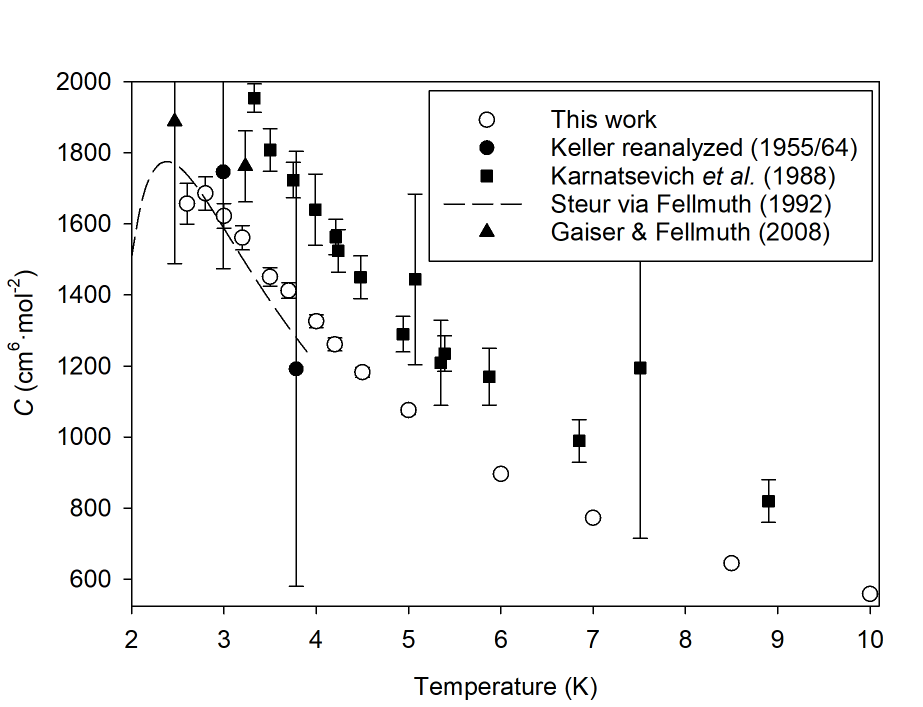}
\caption{The third virial coefficient of ${}^3$He.}
\label{fig:CHe3}  
\end{figure}

There are only a few sources of experimental data for $C(T)$ for ${}^3$He.
Keller~\cite{Keller55} measured five pressure-volume isotherms at temperatures
below 4~K; these were later reanalyzed by Roberts {\em et
  al.}~\cite{Roberts64} and meaningful values of $C$ were obtained only for
the two highest temperatures.  A later analysis of the Keller data was
performed by Steur (unpublished), whose equation for temperatures below 3.8~K
was reported by Fellmuth and Schuster~\cite{Fellmuth92}.  
Some points were also extracted from volumetric data by Karnatsevich {\em et
  al.}~\cite{Karna88} Recently, Gaiser and Fellmuth~\cite{Gaiser08,Gaiser_pc}
extracted virial coefficients from their measurements of two isotherms for
${}^3$He with dielectric-constant gas thermometry.

Figure~\ref{fig:CHe3} compares our calculated values to the available
experimental data, where the error bars represent expanded
uncertainties with coverage factor $k=2$.  Error bars are not drawn
for our values above 5~K because they would be smaller than the size
of the symbol.  As was the case in our previous
work,~\cite{Garberoglio2009b} the uncertainty of our values of $C(T)$
is determined by the statistical uncertainty of our Monte Carlo
calculations (shown in Tables~\ref{tab:CHe4} and \ref{tab:CHe3}) and
by the uncertainty in the two- and three-body potentials. At the
temperatures considered here, the statistical uncertainty is the
dominant contribution to the overall uncertainty. The full uncertainty
analysis is presented elsewhere.~\cite{paper2}

For the experimental points, these expanded uncertainties were taken as
reported in the original sources; we note that in some cases (notably
Ref.~\onlinecite{Karna88}) this appears to be merely the scatter of a fit and
therefore underestimates the total uncertainty.  

Our results are qualitatively similar to the rather scattered
experimental data.  We are quantitatively consistent with the values
based on analysis of the data of Keller, but values from the other
experimental sources are more positive than our results.  We note that
a similar comparison for ${}^4$He~\cite{paper2}, where the
experimental data situation is much better, shows the $C(T)$ values of
Ref.~\onlinecite{Karna88} for ${}^4$He to deviate in a very similar
way not only from our results but from other experimental data we
consider to be reliable.

\section{Conclusions}

We used path-integral methods to derive an expression for the third
virial coefficient of monatomic gases, including the effect of
quantum statistics. We applied this formalism to the case of helium
isotopes, using state-of-the-art two- and three-body potentials.

We showed that exchange effects make no significant contribution
to the third virial coefficient above a temperature of approximately
7~K for both the fermionic and bosonic isotope. This is the same
behavior observed in the calculation of the second virial coefficient.
For temperatures lower than $7$~K, the sign of the contribution to
$C(T)$ from exchange effects depends on the bosonic or fermionic
nature of the atom.
In the case of ${}^4$He, the exchange contribution to $C(T)$ increases
its value compared to the value obtained with Boltzmann
statistics, although in our simulations the total exchange
contribution has the same order of magnitude as the statistical
uncertainty of the PIMC integration.  In the case of ${}^3$He, the
exchange contribution is negative, and its magnitude is much larger
than the statistical uncertainty.

The range of temperatures that we have investigated covers the
low-temperature maximum of $C(T)$ for both isotopes. The third virial
coefficient of ${}^4$He reaches its maximum close to $4$~K, whereas in
the case of ${}^3$He the maximum is attained at a lower temperature.

For both helium isotopes, the uncertainty in our calculated third
virial coefficients is much smaller than that of the limited and
sometimes inconsistent experimental data.  For ${}^4$He, we obtain good
agreement with the most recent experimental results, except for
some temperatures below 10~K.  A full comparison with available
experimental data for ${}^4$He, including the higher temperatures of
importance for metrology, will be presented elsewhere.\cite{paper2} For
${}^3$He, we are qualitatively consistent with the sparse and scattered
experimental values; in this case especially our calculations provide
results that are much less uncertain than experiment.  In both cases,
at the temperatures considered here, the uncertainty is dominated by
the statistical uncertainty of the Monte Carlo integration, meaning
that the uncertainty of $C(T)$ could be reduced somewhat with greater
expenditure of computer resources.

We note two directions in which extension of the present work could be
fruitful.  One is the calculation of higher-order virial coefficients,
which is a straightforward extension of the method presented here.
This would be much more computationally demanding, but the fourth
virial coefficient $D(T)$ may be feasible, at least at higher
temperatures where the number of beads in the ring polymers would not
be large.  Second, the method can be extended to calculate temperature
derivatives such as $\mathrm dC/ \mathrm dT$; such derivatives are of
interest in interpreting acoustic measurements.  Work on the
evaluation of acoustic virial coefficients is in
progress.~\cite{paper2}

\begin{acknowledgments}
We thank C. Gaiser for providing information on low-temperature data
for $C(T)$ of helium isotopes, and M. R. Moldover and J. B. Mehl for
helpful discussions on various aspects of this work.  The calculations
were performed on the KORE computing cluster at Fondazione Bruno
Kessler.
\end{acknowledgments}

\clearpage

\bibliographystyle{aipnum4-1}
\bibliography{virial-He}

\end{document}